\documentclass[aps,prl,superscriptaddress,twocolumn,amsmath,amssymb,longbibliography]{revtex4}
\pdfoutput=1
\usepackage{color}
\usepackage{amsmath}
\usepackage{amssymb}
\usepackage{pdfpages}
\usepackage{bm}
\usepackage{graphicx}
\usepackage{hyperref}
\begin{document}

\title{Polaritons in layered 2D materials}

\author{Tony Low}
\email{tlow@umn.edu}
\affiliation{Department of Electrical \& Computer Engineering, University of Minnesota, Minneapolis, MN 55455, USA}

\author{Andrey Chaves}
\affiliation{Universidade Federal do Cear\'a, Departamento de F\'isica, Caixa Postal 6030, 60455-760 Fortaleza, Cear\'a, Brazil}
\affiliation{Department of Chemistry, Columbia University, New York, New York 10027, USA}

\author{Joshua D. Caldwell}
\affiliation{US Naval Research Laboratory, 4555 Overlook Avenue SW, Washington DC 20375, USA}

\author{Anshuman Kumar}
\affiliation{Department of Electrical \& Computer Engineering, University of Minnesota, Minneapolis, MN 55455, USA}
\affiliation{Mechanical Engineering Department, Massachusetts Institute of Technology, Cambridge, MA 02139, USA}

\author{Nicholas X. Fang}
\affiliation{Mechanical Engineering Department, Massachusetts Institute of Technology, Cambridge, MA 02139, USA}

\author{Phaedon Avouris}
\affiliation{IBM T.J. Watson Research Center, 1101 Kitchawan Rd, Yorktown Heights, NY 10598, USA}

\author{Tony F. Heinz}
\affiliation{Department of Applied Physics, Stanford University, Stanford, CA 94305, USA}

\author{Francisco Guinea}
\affiliation{IMDEA Nanociencia, Calle de Faraday 9, E-28049 Madrid, Spain}
\affiliation{Department of Physics and Astronomy, University of Manchester,
Oxford Road, Manchester M13 9PL, United Kingdom}

\author{Luis Martin-Moreno}
\affiliation{Instituto de Ciencia de Materiales de Aragon and Departamento de Fisica de la Materia Condensada, CSIC-Universidad de Zaragoza, E-50012 Zaragoza, Spain}

\author{Frank Koppens}
\affiliation{ICFO-Institut de Ciencies Fotoniques, The Barcelona Institute of Science and Technology, 08860 Castelldefels (Barcelona), Spain}
\affiliation{ICREA – Instituci$\acute{o}$ Catalana de Recer\c{c}a i Estudis Avancats, 08010 Barcelona, Spain.}





\begin{abstract}
\textbf{In recent years, enhanced light-matter interactions through a plethora of dipole-type polaritonic excitations have been observed in two-dimensional (2D) layered materials. In graphene, electrically tunable and highly confined plasmon-polaritons were predicted\cite{falkovsky2007optical,jablan2009plasmonics} and observed\cite{ju2011graphene,fei2012gate,chen2012optical,yan2013damping}, opening up opportunities for optoelectronics\cite{freitag2013photocurrent}, bio-sensing\cite{rodrigo2015mid} and other mid-infrared applications\cite{low2014graphene}. In hexagonal boron nitride (hBN), low-loss infrared-active phonon-polaritons exhibit hyperbolic behavior for some frequencies\cite{dai2014tunable,caldwell2014sub,caldwell2016atomic}, allowing for ray-like propagation exhibiting high quality factors and hyperlensing effects\cite{li2015hyperbolic,dai2015subdiffractional}. In transition metal dichalcogenides (TMDs), reduced screening in the 2D limit leads to optically prominent excitons with large binding energy\cite{mak2012control,xu2014spin}, with these polaritonic modes having been recently observed with scanning near field optical microscopy (SNOM)\cite{fei2016nano}. Here, we review recent progress in state-of-the-art experiments, survey the vast library of  polaritonic modes in 2D materials, their optical spectral properties, figures-of-merit and application space. Taken together, the emerging field of 2D material polaritonics and their hybrids\cite{caldwell2016atomic} provide enticing avenues for manipulating light-matter interactions across the visible, infrared to terahertz spectral ranges, with new optical control beyond what can be achieved using traditional bulk materials.  }
\end{abstract}

\maketitle

In many materials, electric dipoles (e.g. infrared-active optical phonons, excitons in semiconductors and plasmons in doped materials) can be excited when illuminated\cite{ritchie1957plasma,pekar1958theory}, producing hybrid quasiparticles with photons called \emph{polaritons}. These polaritons can be sustained as electromagnetic modes at the interface between a positive (e.g. normal dielectric) and negative permittivity material, leading to propagating polaritons. In the case of the plasmon-polaritons (PP), the negative permittivity is provided by the coherent oscillations of the free carriers, and can be described by the Drude model. For exciton-polaritons (EP) and phonon-polaritons (PhP), it is associated with their resonant optical absorption, resulting from a highly dispersive permittivity. These optical resonances can also result in a negative permittivity, albeit over a narrow spectral window. In order to provide a simple, yet comprehensive overview of these different optical modes, we have provided a summary in Box 1.

These polariton modes are characterized by two related length scales: the polariton wavelength along the interface and the extension of the evanescent field in the perpendicular direction, both of which are smaller than the free-space wavelength. The associated reduced modal volume presents an extremely large local density of electromagnetic states at the interface, leading to strong light-matter interactions. Hence, polaritonics provide a way to confine, harness and manipulate light at dimensions smaller than the diffraction limit\cite{maier2007plasmonics}. The emergence of so-called 2D or van der Waals materials as hosts for these polaritonic modes, has enabled the first realization and imaging of polaritons within atomically thin materials.  Due to the inherent material anisotropy and the exceptional variation of material types (e.g. metallic, dielectric, semiconductor) available within the known library of 2D materials, a large breadth of different flavors of polaritonic modes have been realized. This includes tunable chiral\cite{song2015chiral,kumar2016chiral}, anisotropic and hyperbolic PPs\cite{low2014plasmons,nemilentsau2016anisotropic}; long-lived hyperbolic PhPs\cite{dai2014tunable,caldwell2014sub,caldwell2015low} with slow light\cite{yoxall2015direct,caldwell2015mid} and hyperlensing behavior\cite{li2015hyperbolic,dai2015subdiffractional}; EP with strong binding energies\cite{splendiani2010emerging,mak2010atomically,conley2013bandgap,ross2013electrical,ugeda2014giant}, their complexes\cite{mak2013tightly,ross2013electrical,jones2013optical,you2015observation}, 
and anisotropic excitons\cite{zhang2014extraordinary,castellanos2014isolation,yang2015optical,wang2015highly}. This review is devoted to the emerging but rapidly developing field of polaritonics and nanophotonics in the family of 2D materials, with an emphasis on the materials and the polaritons physics.  

\emph{From metal to graphene plasmon-polaritons.} 
The most well-known physical realization of polaritons consists of electromagnetic modes bound to a flat interface between a metal and a dielectric, called surface plasmon-polaritons (SPPs)\cite{maier2007plasmonics}. The field of metal plasmonics has developed  tremendously over the last few decades, with a number of interesting effects such as extraordinary transmission through nanoholes in metals\cite{ebbesen1998extraordinary}, single molecule detection\cite{nie1997probing}, compact nanophotonic components\cite{ozbay2006plasmonics}, and novel optical phenomena with metamaterials\cite{shalaev2007optical} and metasurfaces\cite{yu2014flat}. However, metal plasmonics suffers from the problem of absorption losses\cite{khurgin2015deal}, which traditionally has limited the range of possible materials to metals like silver and gold, and constrains the operating frequencies to the near-infrared, visible and ultraviolet. 

\emph{State-of-the-art graphene plasmonics.} 
Graphene plasmonics\cite{falkovsky2007optical,jablan2009plasmonics,koppens2011graphene,nikitin2011fields,low2014graphene} presents several advantages as compared to metal plasmonics. First, the carrier density, which determines its plasmonic Drude weight (as defined in Box 2), can be electrically\cite{ju2011graphene}, chemically\cite{yan2013damping}, or optically\cite{ni2016ultrafast} tuned. This is due to the fact that graphene is a semimetal with a small density-of-states, where typical carrier concentrations are less than $\sim 0.01$ free conduction electrons per atom, in contrast to the case of $\sim 1$ in gold. The consequence of this lower carrier concentration is that graphene PPs can be stimulated in the terahertz to mid-infrared frequencies\cite{ju2011graphene,fei2012gate,chen2012optical,yan2013damping,low2014graphene},
and in some cases even into the short-wave infrared\cite{tielrooij2015electrical}, while
metal SPPs are found in the UV, visible to the near-infrared. Second, the low electronic density-of-states and relatively weak electron-phonon coupling endows graphene with very high intrinsic carrier mobilities, attainable via graphene encapsulated within hexagonal boron nitride (hBN) multilayers made with van der Waals assembly technique\cite{wang2013one}.

First experiments\cite{fei2012gate,chen2012optical} on propagating PPs in graphene were observed using scattering-type scanning near field optical microscopy (SNOM). Recently, similar measurements have been performed on high mobility encapsulated graphene\cite{woessner2015highly}, with a schematic of the setup presented in Fig.\,1a. In this experiment, the metal-coated AFM tip scatters incident free-space light into graphene PPs, where the sharpness of the tip provides the needed momentum to overcome the momentum mismatch between free-space photons and the confined PPs. The PP propagates away from the tip as circular waves with complex wavevector $q$. 
The PP propagates towards the edge of the graphene flake, and provided this edge is within the polariton propagation length, it will be reflected back towards the tip and is detected as out-scattered light. Spatial scanning of the tip near the graphene edge shows characteristic fringes due to interference between the reflected and incident PPs (see Fig.\,1b) from which the complex $q$ can be extracted. In such experiments, spacing between fringes is the PP half-wavelength i.e. $\pi/\mbox{Re}[q]$, while the fringes decay exponentially as $\mbox{exp}(-\mbox{Im}[q]x)$. 

In order to directly compare and quantify the behavior of various polaritonic materials, a series of figures of merit have been established. Commonly used figures-of-merit for PPs are; $\gamma\equiv \mbox{Im}[q]/\mbox{Re}[q]$, where $\gamma^{-1}/2\pi$ gives the number of cycles the PP can propagate before its amplitude decays by $1/e$, and the light confinement factor $\beta$, which is obtained by normalizing $\mbox{Re}[q]$ to the free-space wavevector. Experiments with hBN encapsulated graphene have achieved $\beta\sim 150$ and $\gamma^{-1}> 25$\cite{woessner2015highly}, see Fig.\,1c. Hence, the current state-of-the-art in graphene PP has already surpassed performance of SPPs along air-silver interfaces ($\gamma^{-1}\approx 10$ across the visible spectral range, where $\beta>10$\cite{jablan2009plasmonics}). In these graphene-based devices, the inverse damping $\gamma^{-1}$ was found to be dominated by phonons in graphene and hBN\cite{principi2014plasmon}, rather than that of extrinsic ionized impurities as in earlier experiments on SiO$_2$ substrates\cite{fei2012gate,chen2012optical} where $\gamma^{-1}\sim 5$.  Steps have also been undertaken to develop resonant optical gold rod antennas and 2D spatial conductivity patterns for direct launching and control of propagating PPs\cite{alonso2014controlling} as illustrated in Fig.\,1d and e respectively.

\emph{Plasmon-polaritons beyond graphene.} Graphene has paved the way for the discovery and exploration of other atomic 2D materials with new physical properties\cite{novoselov2005two}, among them new types of PPs. For example, Bernal-stacked bilayer graphene has recently received attention as a new plasmonic material, both theoretically\cite{low2014novel} and experimentally\cite{fei2015tunneling,yan2014tunable}. It allows for efficient switching of the plasmonic Drude weight with the application of a vertical electric field\cite{fei2015tunneling}, due to the opening of an electronic band gap\cite{oostinga2008gate}. It also accommodates a mid-infrared-active optical phonon mode\cite{kuzmenko2009gate}, which can hybridize with the PPs to feature novel effects such as electromagnetically induced transparency and slow light\cite{yan2014tunable}. This, in addition to other new emerging 2D materials such as TMDs\cite{mak2012control,xu2014spin} and black phosphorus (BP)\cite{li2014black}, presents exciting opportunities for exploring new plasmonic effects\cite{low2014plasmons,song2015chiral,kumar2016chiral,nemilentsau2016anisotropic} (Box 2). 

As gold has perhaps been the most utilized plasmonic material, it is instructive to compare the associated figures-of-merit for the corresponding SPPs with those of 2D materials. In the mid-infrared, due to the large negative values associated with the real part of the dielectric function, SPPs in gold exhibit very poor $\beta$. However, $\beta$ can be increased by considering metal films, and utilizing the ``short-range SPP'' mode (Box 1). To facilitate comparison, we consider the extreme case (most optimal in terms of $\beta$) assuming a $0.235\,$nm thick gold film (equal to the separation between (111) atomic planes). Bulk dielectric constants for gold are used\cite{johnson1972optical}. Optical constants for graphene and its bilayer, MoS$_2$, and BP have been obtained from the well-known Kubo formula in conjunction with their low-energy Hamiltonians\cite{low2014novel,low2014plasmons,kumar2016chiral}. In all cases, we have assumed a temperature of $300\,$K and a n-doped material with a typical doping of $5\times10^{12}\,$cm$^{-2}$ which can be obtained with standard electrical gating\cite{novoselov2005two}. We obtained the free carrier scattering time $\tau$ from current state-of-the-art experiments\cite{woessner2015highly,li2015quantum,cui2015multi}. For graphene, its bilayer, and MoS$_2$, $\tau$ is approaching the intrinsic limit determined by their thermal and optical phonons, i.e. $\tau\sim 1\,$ps\cite{principi2014plasmon}  and $\sim 0.1\,$ps\cite{kaasbjerg2012phonon} respectively. For BP, $\tau$ can approach $\sim 1\,$ps\cite{qiao2014high}, however, the best experiments to-date achieved only $\sim 0.1\,$ps\cite{li2015quantum}, so for our purposes the latter value was used. All calculations are for a configuration where the 2D material is placed on a substrate with dielectric constant of $2.25$, which corresponds to that of SiO$_2$ in the mid-infrared.

Fig.\,1f and g show the calculated field confinement $\beta$ and damping ratio $\gamma^{-1}$ across the far- and mid-infrared frequencies for 2D plasmonic materials and the fictional atomic gold film. In all cases, $\beta$ is observed to increase with increasing frequency, however, it is only of order $1$ for the gold film even upon approaching the near-infrared. On the other hand, 2D materials present much larger $\beta\sim 100-1000$. This implies 2D materials are clearly better than metals for field enhancement and localization in the mid-infrared. From these plots, it is clear that no one material provides the optimal behavior at all frequencies.  For instance, while the highest $\beta$ values are observed in BP, these are observed along a single crystallographic axis ($x$) and only over a relatively narrow spectral window.  While this inherent in-plane anisotropy gives rise to the potential for hyperbolic PPs (Box 2), this may not be ideal for many applications, especially when identifying the crystallographic facets within micron-sized flakes can be highly problematic. In contrast, graphene and MoS$_2$ both provide high $\beta$ over a broad spectral range extending from the terahertz out into the mid- and near-infrared, respectively, with MoS$_2$ offering higher $\beta$ values overall.  In the case of graphene, the operational range is limited on the high-frequency end to about $3500\,$cm$^{-1}$ due to interband losses.  These occur when the plasmon energy is roughly twice the Fermi energy\cite{wunsch2006dynamical}.  A very interesting behavior is observed in bilayer graphene.  Within the mid-IR, $\beta$ is found to degrade rapidly with increasing frequency.  However, after an extended spectral gap between roughly $1450-3250\,$cm$^{-1}$, the $\beta$ recovers.  This behavior is related to the two nested conduction bands in the bilayer, which due to interlayer coupling can transfer the energy to another higher energy plasmonic mode\cite{low2014novel}.  Thus, bilayer graphene can be used to extend polaritonics into the near-infrared.

Semiconducting TMDs, especially monolayer MoS$_2$, have attracted significant attention due to their novel optical dichroic and coupled spin-valley physics \cite{xu2014spin}. Despite the considerable bandgap of $\sim 2\,$eV, a metallic state can still be induced with a vertical electric field if gap trap-states\cite{zhu2014electronic} can be sufficiently suppressed. The development of van der Waals heterostructure device platforms where MoS$_2$ layers are fully encapsulated within hBN represents an important progress in this direction, with a record-high Hall mobility reaching $34,000\,$cm$^2$V$^{-1}$s$^{-1}$ for six-layer MoS$_2$ at low temperature\cite{cui2015multi}. Fig.\,1f shows that TMDs can have a very large confinement of $\beta\sim 10^3$, an order larger than graphene, albeit its $\gamma^{-1}$ is smaller over most of the spectral range where graphene plasmons can be supported due to it smaller carrier lifetime. A new class of anisotropic 2D materials\cite{li2014black,tongay2014monolayer,island2014ultrahigh} has also received considerable attention recently, particularly BP, which has a bulk gap of $0.3\,$eV and decent carrier mobility of $\sim 1000\,$cm$^2$V$^{-1}$s$^{-1}$ for BP thin film\cite{li2014black}. In BP, the in-plane electronic mass anisotropy can be as large as $10$, which as discussed above offers the promise of anisotropic PPs\cite{low2014plasmons} as reflected in their $\beta$ and $\gamma^{-1}$ presented in Fig.\,1f and g. 

\emph{Hyperbolic phonon-polaritons in hBN.}
While the in-plane anisotropy of BP provides the potential to realize hyperbolic PPs behavior within the plane of a 2D material (e.g. Box 2), the structural anisotropy in all van der Waals crystals results in a strong optical birefringence. Hyperbolicity is defined as an extreme type of birefringence, whereby the permittivities along orthogonal crystal axes are not just different, but opposite in sign\cite{poddubny2013hyperbolic}. In 2D materials, the Cartesian axes  exhibiting the sign inversion of the permittivity tensors tend to be between the in- and out-of-plane directions, with hBN offering a prime example. Bulk hBN features two sets of infrared active optical phonon modes, which results in two spectrally distinct Reststrahlen bands where negative permittivity can be observed.  These two bands are
 now referred to as the lower (``LR''; $\sim 760-825$cm$^{-1}$) and upper (``UR''; $\sim 1360-1620$cm$^{-1}$) Reststrahlen bands. As shown in Fig.\,2a, its in-plane permittivity is positive (negative) within the LR (UR), while opposite in sign along the out-of-plane $c$-axis.  

The inversion of signs of the permittivity components between these two bands gives rise to a type I and II hyperbolic behavior within the LR and UR respectively\cite{poddubny2013hyperbolic}, which impacts the resultant dispersion relationships of the in-plane transverse ($k_t$) and out-of-plane ($k_z$) hyperbolic modes. 
These hyperbolic PhPs in hBN were first explored within the UR using SNOM\cite{dai2014tunable} through a similar interferometric study to that previously described for PP in graphene. This initial experiment recorded $\beta\sim 25$ and $\gamma^{-1}\sim 20$, which further improved to 
$\gamma^{-1}\sim 35$ in a subsequent experiment\cite{dai2015subdiffractional}. Such high $\gamma^{-1}$ is not surprising since PhPs do not suffer from electronic losses, instead being dictated by the optic phonon scattering lifetimes, which are intricately tied to the crystal structural quality and should improve with better growth.

Unlike traditional polaritonic materials, hyperbolic PhP modes exhibit multiple branches within the dispersion relationship as shown in Fig.\,2b. 
 Later work demonstrated that these multiple branches are directly tied to the thickness of the hyperbolic material layer thickness\cite{dai2015subdiffractional}.
While the initial work\cite{dai2014tunable} observed only the fundamental, lowest order hyperbolic modes, work using conical-shaped hBN nanostructures experimentally observed up to 7 and 4 branches within the UR and LR, respectively, by varying the aspect ratio of the structures\cite{caldwell2014sub} (Fig.\,2c). 
Unlike the case of SNOM, where the higher order branches exhibit higher in-plane wavevector, $k_t$, in the resonators, the modes increased in $k_z$.  As the aspect ratio for these nanostructures is directly proportional to the out-of-plane $k_z$, this gives rise to the apparent inversion of the dispersion relationships between Fig.\,2b (SNOM) and 2c (nanostructures). Subsequent work using SNOM was able to further map out the first few orders of the $k_t$ dispersion as well\cite{dai2015subdiffractional}.

\emph{Slowing light and imaging within hBN.} The PhP dispersions reported in Ref. \cite{dai2015subdiffractional} and \cite{caldwell2014sub} suggest negative (positive) group velocity for the $k_z$ ($k_t$) modes within the UR, and the opposite sign for the LR.  Interestingly, recent time-resolved SNOM measurements provided unambiguous evidence that it is instead the phase velocity which exhibits negative values\cite{yoxall2015direct,caldwell2015mid}. Fig.\,2d depicts the launching of PhP from the edge of a gold pad, and visualizing its propagation in the time-domain for PhP in the UR (Fig.\,2e) and LR bands (Fig.\,2f).  By tracing its wave envelope, one can discern that the PhP moves with a positive group velocity 
regardless of the Reststrahlen band. However, by monitoring the fringe velocity, the sign of the phase velocity can be identified as being negative in the LR, while positive in the UR.  Such a negative phase velocity necessarily implies that $k_t$ is negative. Presumably, for modes where $k_z$ is changing (e.g. 3D confined resonators), the opposite behavior would be anticipated. These experiments also allowed for direct measurement of the hyperbolic PhP modal lifetimes, recorded to be $1.8$ and $0.8\,$ps in the LR and UR, respectively, demonstrating that these values are on the order of the lifetimes of the optic phonons themselves. With lifetimes for PPs in metals being on the order of $10$s fs, such long lifetimes would naturally be anticipated to result in correspondingly long polariton propagation lengths.  However, as previously discussed\cite{caldwell2015low}, the group velocity at which the polariton propagates can be exceptionally slow, therefore limiting the ultimate propagation length.  For hBN, this was certainly the case with the group velocity being demonstrated to be on the order of $0.027$ and $0.002$ times the speed of light in vacuum for the UR and LR, respectively.

Another promising application for hyperbolic media that has received a lot of attention is the direct imaging of deeply sub-diffractional sized objects via the so-called ``hyperlens''\cite{jacob2006optical}.   Originally demonstrated using hyperbolic metamaterials\cite{liu2007far} in the visible, the high optical losses associated with the metal plasmonic constituents and the fabrication complexities have limited its imaging capabilities and resolution to the smallest meta-atom. On the contrary, hyperbolic materials based on hBN are homogeneous, ideally infinite, and low-loss by nature of PhP. Proof-of-principle hyperlens experiments were recently demonstrated\cite{li2015hyperbolic,dai2015subdiffractional}, as illustrated in Fig.\,2g-i. In both works, sub-diffractional metal objects were fabricated on a substrate surface using electron beam lithography (Fig.\,2h) and covered with a flat slab of hBN (on the order of $100\,$nm thick) and imaged using a SNOM (Fig.\,2g). Multiple rings result from the PhP modes launched from the edges of the metal nanoparticles at  angles dictated by the hyperbolic nature of the PhPs(Fig.\,2i). This propagation angle (with respect to the surface normal) results from the fact that while hyperbolic materials can support very high $k$ modes, they can only propagate at an angle given by $\mbox{tan}(\theta)=\sqrt{\mbox{Re}[\epsilon_t]/\mbox{Re}[\epsilon_z]}/i$. For both the UR and LR of hBN, either $\epsilon_t$ or $\epsilon_z$ will be negative and dispersive while the other is positive and nominally constant, this angle of propagation is therefore a function of frequency.  Because of this, the image that results from the corresponding hyperlens can vary from a direct replica ($90^o$ propagation) to a highly magnified object, which can be user-controlled. 
It is important to note that the propagation angle $\theta$ is fixed with respect to the crystal axes of hBN. As a result, in a non-planar geometry of a truncated cone geometry (as illustrated in the inset of Fig.\,2c), non-specular reflections can occur off the sidewall surfaces as recently confirmed with SNOM\cite{giles2016imaging}.

\emph{Natural hyperbolic layered 2D materials beyond hBN.} To this point, research in this area has focused on the natural hyperbolic properties of hBN. However, the natural optical anisotropy associated with van der Waals crystals, and the polar nature of many should in principle offer a broad range of naturally occurring hyperbolic materials covering a very broad spectral range\cite{sun2014indefinite,korzeb2015compendium,narimanov2015metamaterials}. 
Strong anisotropy in electron motion along the in-plane (metallic) and out-of-plane (insulating) layered materials can lead to hyperbolicity for specific frequency bands, e.g. in graphite and magnesium diboride\cite{sun2011indefinite,sun2014indefinite}. Ruthenates  have different Drude weights\cite{sun2014indefinite} along the in- and out-of-plane axes, and are hyperbolic between the two plasma frequencies. High quality semiconducting TMDs, such as MoS$_2$, also accommodate far-infrared anisotropic polar optical phonons, should result in hyperbolic bands as in hBN in the Reststrahlen frequencies\cite{wieting1971infrared}. Furthermore, the tetradymites, such as Bi$_2$Se$_3$ and Bi$_2$Te$_3$, should also host such hyperbolic polaritons, albeit in the terahertz regime\cite{wu2015topological} , while they were also observed to support hyperbolic modes in the near-infrared to visible due to different highly resonant interband transition energies along the in-plane and out-of-plane axes\cite{esslinger2014tetradymites}. Lastly, the high T$_c$ superconducting cuprates are highly metallic in-plane, but with an out-of-plane dielectric response typical of an insulator characterized by several Lorentzian type resonances\cite{sun2014indefinite}.  As such, one would naturally infer that hyperbolic behavior would also be readily observed. 

Fig.\,2j presents representative materials from the above-mentioned layered 2D systems. Their hyperbolicity spectral range, type, and inverse damping ratio $\gamma^{-1}$ are plotted. Here, we have used the quasistatic approximation and considered only the lowest order mode. This allows us to characterize a broad class of materials on  equal footing. Among the listed materials, only hBN and Bi$_2$Se$_3$ whose hyperbolicity are phonon in origin, show $\gamma^{-1}>10$, indicating that while a wealth of hyperbolic 2D materials await further study, optical losses may still limit advancements in the near-term.

\emph{Strong excitons in 2D semiconductors.} Over the past few years, a plethora of photoluminescence (PL), absorption, and reflectance experiments have been reported in the literature for several 2D materials, such as TMD monolayers\cite{splendiani2010emerging,mak2010atomically,conley2013bandgap,ross2013electrical,ugeda2014giant} (including the distorted 1T phase ReS$_2$\cite{aslan2015linearly}), BP mono and multilayers\cite{wang2015highly,zhang2014extraordinary,castellanos2014isolation,yang2015optical}, and, more recently, monolayer organic-inorganic perovskite (OIP) crystals\cite{yaffe2015excitons}, namely (C$_4$H$_9$NH$_3$)$_2$PbI$_4$. By probing the formation (absorption and reflectance) and recombination (PL) of electron-hole pairs, these experiments have provided us information not only about their optical gaps, which are found to be significantly far from the theoretical quasi-particle bandgaps, but also about their excitonic Rydberg series. Figure 3a summarizes the experimentally observed optical gaps (full symbols), which cover a wide range of wavelengths from blue, for monolayer OIPs, all the way to the near-infrared, for BP multilayers. Recent calculations also suggests possible extension into the mid-infrared using arsenene, a stable analogue of BP but with arsenic atoms\cite{chaves2016theoretical}. 

Binding energies of excitonic states can be obtained by $E_n = E_{Xn} - E_{QP}$, where $E_{Xn}$ and $E_{QP}$ are the $n$-th exciton state energy and quasi-particle band gap, respectively, and are shown in Fig.\,3b. Unlike bulk semiconductors, excitons in 2D materials are strongly confined to a plane and experience a reduced screening from their surrounding dielectric enviroment\cite{berkelbach2013theory}, which modifies the character of the Coulomb interaction potential, leading to non-hydrogenic Rydberg series of exciton states\cite{chernikov2014exciton}. Excited exciton states within the Rydberg series  (i.e. $2s$, $3s$, $4s$... states) have been observed in OIPs\cite{yaffe2015excitons} and TMDs\cite{chernikov2014exciton,he2014tightly,ye2014probing,zhu2015exciton}, using different methods such as reflectance and absorption spectroscopy, as summarized by circles in Fig.\,3b. Triangles in Fig.\,3b are excited states with $p$ symmetry, observed by two-photon luminescence experiments in WS$_2$ and WSe$_2$\cite{he2014tightly, ye2014probing, zhu2015exciton}. As a consequence of the non-Coulombic electron-hole interaction potential, the degeneracy between $s$ and $p$ states is lifted, so that two-photon experiments exhibit peaks in the PL spectra that do not match those of the Rydberg series states\cite{ye2014probing}. In general, binding energies in 2D materials are about an order of magnitude higher than those of bulk semiconductors,\cite{koch2006semiconductor} such as Si, Ge and III-V or II-VI alloys, being comparable only to exciton binding energies observed in carbon nanotubes\cite{wang2005optical} and conjugated polymer chains \cite{sebastian1981one}. 

Such strong electron-hole interaction makes these materials a playground for investigating excitons and their complexes (trions and biexcitons), which are usually harder to observe in conventional bulk semiconductors. In Fig.\,3b, we summarize the recently experimentally observed trions and biexcitons binding energies in 2D materials, which ranges from tens to a hundred meV\cite{jones2013optical,mak2013tightly,mai2013many,ross2013electrical,plechinger2015identification,zhang2015valence,you2015observation,xu2016extraordinarily}. 
All these high binding energies for excitons and their complexes as observed are consistent with theory within the generalized Wannier-Mott model described elsewhere\cite{chaves2015anisotropic}. 

In tightly bound excitons in layered 2D materials, there is a large spatial overlap of the respective electron- and hole-wavefunctions, with the corresponding Bohr radius being on the order of only $1\,$nm\cite{chernikov2014exciton,ye2014probing}. This leads to a particularly strong coupling of excitons to photons\cite{haug1990quantum}, resulting in both large absorption coefficients and efficient emission of the radiation in this class of materials. The absolute absorption values at the excitonic transition peak are as high as $10-20\%$ \cite{mak2010atomically} for single layers with sub-nanometer thickness, as shown in Fig.\,3c. The corresponding area under the resonance (shaded), proportional to the strength of the light-matter coupling\cite{klingshirn2012semiconductor}, is orders of magnitude larger than the respective values in more conventional inorganic semiconductors, such as GaAs\cite{masselink1985absorption}.

\emph{Exciton-polaritons.} The EP was first observed in 2D materials within optical microcavities with TMDs\cite{gan2013controlling,liu2015strong}. Experimentally observed EP branches\cite{liu2015strong} in monolayer MoS$_2$ follow anti-crossing paths over the exciton and cavity mode energy lines, with a Rabi splitting as large as $\sim 46\,$meV. The strong coupling regime, typically defined by the rate of the exciton-phonon scattering being larger than the competing dephasing processes of the two particles, was already shown to be attainable even at room temperature\citep{liu2015strong}. However, real-space observation of EP with SNOM was observed only very recently\cite{fei2016nano} in an exfoliated $260\,$nm WSe$_2$ thin flake. The field of EPs in 2D materials is still at the nascent stage, and we expect exciting future developments.

\emph{Hybrid polaritons.} Looking forward, perhaps one of the most intriguing prospects associated with 2D van der Waals crystals is the ability to cleave and combine layers of different 2D materials to realize heterostructures of different constituents and thicknesses\cite{geim2013van}, and engineer new hybrid polaritons with novel physics\cite{fang2014strong,Fogler2014,Rivera2015,caldwell2016atomic}. Two recent examples are depicted in Fig.\,4. By combining graphene with hBN, one can marry the advantage of tunable PP in graphene with high quality, low-loss, PhP in hyperbolic hBN\cite{dai2015graphene}. Recent SNOM measurement performed on a graphene-hBN heterostructure\cite{dai2015graphene} (Fig.\,4a) revealed a coupled PP and PhP  (Fig.\,4b). This hybrid polariton mode, termed a 'plasmon-phonon polariton', exhibited a broad-band dispersion, extending beyond the Reststrahlen band of hBN, and offered electrostatic gate tuning of the hyperbolic mode primarily confined within the hBN.  In addition to these so-called electromagnetic hybrids where combinations of PPs and PhPs are induced, another  new type of atomic-scale heterostructure consisting of polar dielectric 2D materials can also lead to the creation of a new material resulting from the hybridization of optic phonons at the heterostructure interfaces, which can provide direct control of the dielectric function within the spectral regime of the Reststrahlen bands\cite{caldwell2016atomic}.

On the EP front, heterostructures of TMDs where electrons and holes are confined to different layers, as illustrated in Fig.\,4c, would allow for the formation of ``indirect excitons'' at an energy lower than that of its single layer constituents. Many pairs of 2D materials are known to be compatible with the type-II band alignment\cite{ozccelik2016band} required for such situations. Having a low oscillator strength, but also lower energy, this charge transfer exciton manifests itself as a clear low energy peak in PL experiments that is absent in absorption experiments\cite{Rivera2015}, as shown in Fig.\,4d for a WSe$_2$/MoS$_2$ hybrid heterostructure\cite{fang2014strong}. 

\emph{Outlook.}
As indicated by the breadth of work highlighted above, there is a great deal of promise for nanophotonics based on 2D materials and their heterostructures.  While still in the earliest stages of work, here we offer our perspective on how we anticipate the development of the field, highlighting the key physics and the potential application space.
The latter is summarized in Box 1. Graphene plasmonics has led the way in terms of demonstrating proof-of-principle device concepts such as mid-infrared optoelectronics\cite{freitag2013photocurrent}, bio-sensing\cite{rodrigo2015mid} and fingerprinting\cite{hu2016far}. We anticipate that these applications should continue to develop over the next few years, and also envision ventures into free-space beam shaping and steering with graphene metasurfaces\cite{carrasco2015gate}. Going forward, these high quality graphene PPs can also provide an excellent platform for realizing tunable 2D mid-infrared nanophotonics circuits with novel functionalities not previously attainable\cite{vakil2011transformation}. 

The discovery of new plasmonic materials beyond graphene might potentially offer new means of light-matter interactions, such as hyperbolic\cite{nemilentsau2016anisotropic} and chiral\cite{kumar2016chiral,song2015chiral} plasmons in anisotropic and gapped Dirac materials respectively. Plasmonic loss remains a fundamental issue that should be addressed, and non-equilibrium processes or gain media\cite{page2015nonequilibrium} could offer enticing routes, particularly in gapped materials e.g. bilayer graphene. However, while there is a strong desire to extending the operating spectral range into the visible, this would also necessarily require the realization of new materials, for instance intercalated graphene\cite{khrapach2012novel} or the recently observed  hyperbolic polaritons in tetradymites\cite{esslinger2014tetradymites}.

Because of the optical phonon origin of PhPs, these modes are observed in the THz to mid-infrared spectral ranges\cite{caldwell2015low}, as summarized in Fig.\,2j. In 2D materials, the natural hyperbolic character of these phononic modes are preferred candidates for sub-diffraction imaging (such as hyperlensing discussed earlier), and super-Planckian thermal emission\cite{guo2012broadband}. However, while the slow moving PhP modes are not ideal for most waveguide applications, the long residence times associated with these long modal lifetimes and slow velocities can be extremely beneficial for enhancements of local emitters, molecular vibrational modes and applications where strong-light matter interactions are desired. Indeed, these slow-light modes in natural hyperbolic materials would also manifest a van Hove singularity in the near-field local density of optical states\cite{cortes2013photonic}, offering control of spontaneous emission.  

Long-lived excitons are of great importance e.g. for possibly allowing future observation of their superfluidity and exciton condensation at relatively high temperatures\cite{Fogler2014}. Such unprecendented tightly bound exciton complexes also brings the exciting perspective of realizing efficient energy transfer by driving charged excitons through applied electric fields. This could be interesting for future applications in solar cells and photodetectors\cite{lee2014atomically}, where suppression of electron-hole recombination is also desireable. The coupling of excitons with plasmons to achieve sustained propagating EPs\cite{fei2016nano} is also interesting. Last but not least, the emergence of topological materials with nontrivial ``twisted'' electronic wavefunctions has also motivated studies on imbuing topological property to excitons\cite{srivastava2015signatures}, plasmons\cite{kumar2016chiral,song2015chiral} and phonons\cite{zhang2015chiral} in solids. This could potentially open up new avenues for the observation of topological polariton physics. 

In summary, the use of 2D polaritons has enabled the engineering of light-matter interactions beyond the diffraction limit across the terahertz to visible spectrum. The ability to manipulate polaritons within the vast library of van der Waals 2D materials, in addition to nano- and heterostructuring, promises the on-demand design of new optical properties which are not possible with traditional plasmonic materials.

\newpage

\textbf{Figure 1. State-of-the-art graphene plasmonics.}
 (a) Schematic of the SNOM measurement, where the probe tip is excited with a laser source, launching plasmons radially from the tip, with the scattered plasmon also collected by the tip.  (b) Measured optical signal from a 2D scan of the SNOM tip near the graphene edge (dashed line) at room temperature. Plasmons are reflected off the graphene edge, and appear as interference fringes. 
(c) Calculated graphene plasmon inverse damping ratios, $\gamma^{-1}$, due to graphene acoustic phonons (blue dashed line), substrate phonons of hBN (yellow dashed line), and the combination of these mechanisms. $\gamma^{-1}$ due to charge impurities at concentration $1.9\times 10^{11}\,$cm$^{-2}$ (green dashed-dotted line) are also displayed. Experimentally measured $\gamma^{-1}$ are shown in solid symbols with error bars representing the $95\,\%$ confidence intervals. (d) Experimental SNOM image of a convex Au antenna extremity due to laser excitation at $11.06\,\mu$m, demonstrating the possibility of plasmon launching and wavefront engineering in graphene. (e) Similar SNOM image of refraction of graphene plasmon launched from Au antenna due to a graphene bilayer prism as indicated. 
Calculated confinement factor, $\beta$, and inverse damping ratio, $\gamma^{-1}$, for various 2D materials such as graphene, bilayer graphene, black phosphorus (BP), and MoS$_2$, displayed in (f) and (g) respectively. BP exhibits highly anisotropic in-plane electronic dispersion, with effective masses along the two crystal axes differing by a factor of $\sim 10$. We displayed results for both the high ($x$) and low ($y$) mass directions. 
(a-c) adapted with permission from Ref.\,\cite{woessner2015highly}. (d-e) adapted with permission from Ref.\,\cite{alonso2014controlling}. 
\\

\textbf{Figure 2. Hyperbolic phonon-polaritons in hBN.}
(a) Real parts of the in-plane ($\epsilon_t$) and out-of-plane ($\epsilon_z$) permittivity tensor components of hBN. Type I lower and II upper Reststrahlen bands are shaded. A schematic of the hBN crystal structure is presented in the inset. 
(b) Phonon-polariton dispersion within Reststrahlen bands experimentally obtained from the SNOM images near edges of a $105\,$nm thick hBN, versus the in-plane momenta $k_t$ (solid symbols). Solid lines are theory results. 
(c) Hyperbolic phonon polariton dispersion of the out-of-plane wavevector $k_z$ as determined by the aspect ratio dependence of the resonance frequencies for difference conical shaped nanostructures.  This is plotted for both the upper (top) and lower (lower) Reststrahlen band.  The solid lines are analytical calculations for ellipsoidal particles, while the various symbols indicate the resonant frequencies for experimental conical nanostructures.
(d) Schematic of time-domain SNOM measurement of phonon-polariton in hBN. Mid-infrared light incident on the Au antenna launches hyperbolic polaritons in hBN, which propagate away from the Au edge and decay exponentially in amplitude and are finally collected by the nanotip. (e-f) Line scans of the SNOM amplitude taken as a function of the delay time between the incident (on Au) and detected fields (by tip). The polaritons group velocity (measured for frequency within the type II and I Reststrahlen bands respectively) can be extracted from the rate at which the `envelope' of the fields propagate away from the Au edge, while both sign and magnitude of the phase velocity can be determined from the direction and speed of the red/blue fringes with respect to the envelope. (g) Schematic showing the launching of hBN hyperbolic phonon polaritons from edges of Au disc, when it is illuminated with mid-infrared light. (h) AFM image of Au disks defined lithographically on SiO$_2$/Si substrate before the hBN transfer. (i) SNOM image of a $395\,$nm thick hBN at laser frequency $\omega=1515\,$cm$^{-1}$, where the observed ``rings'' produced by the hyperbolic polaritons are concentric with the Au disks. 
(j) Chart showing the type I and II hyperbolic frequency ranges for various naturally occuring hyperbolic layered materials, i.e. cuprates\cite{sun2014indefinite} (BSCCO (Bi$_2$Sr$_2$Ca$_{n-1}$CunO$_{2n+4+x}$), GBCO (GdBa$_2$Cu$_3$O$_{7-x}$), LASCO (La$_{1.92}$Sr$_{0.08}$CuO$_4$)), ruthenates\cite{sun2014indefinite} (SRU (Sr$_2$RuO$_4$), SR3U (Sr$_3$Ru$_2$O$_7$)), hBN\cite{dai2014tunable,caldwell2014sub}, TMDs\cite{wieting1971infrared}, tetradymites (Bi$_2$Se$_3$)\cite{wu2015topological}, and graphite\cite{sun2011indefinite,sun2014indefinite}. Color map depicts the calculated figure of merit $\Re\{q\}/\Im\{q\}$ for the lowest polaritonic mode in the quasi-static limit. Dielectric functions of these materials are obtained from references as cited.
(a,c) adapted with permission from Ref.\,\cite{caldwell2014sub}. (b,g-i) adapted with permission from Ref.\,\cite{dai2015subdiffractional}. (d-f) adapted with permission from Ref.\,\cite{yoxall2015direct}.
\\

\textbf{Figure 3. Excitons in TMDs and beyond.} 
(a) Survey of experimentally observed optical band-gaps (full symbols) in different 2D materials: OI in both phases (I and II) \cite{yaffe2015excitons}, WS$_2$ , WSe$_2$ \cite{chernikov2014exciton, he2014tightly, ye2014probing, zhu2015exciton}, MoS$_2$,\cite{conley2013bandgap, splendiani2010emerging, mak2010atomically}, MoSe$_2$ \cite{ugeda2014giant, ross2013electrical},  ReS$_2$ \cite{tongay2014monolayer, aslan2015linearly}, ReSe$_2$ \cite{yang2015tuning}; and $n$-layers black phosphorus (n-BP). \cite{wang2015highly,zhang2014extraordinary,castellanos2014isolation,yang2015optical}.  
We note that for anisotropic materials like ReS$_2$, ReSe$_2$ and n-BP, the optical spectra are polarization sensitive, and the optical gap is defined by the lowest energy resonance.
Open symbols are theoretical calculations based on the Wannier Mott model.  Theoretical predictions of quasi-particle (open circles) \cite{zhang2015manifestation} and optical (blue stars) gaps for $n$-As are shown, extending the frequency window into the mid-infrared. (b) Binding energies of exciton states in upper panel, in their Rydberg sequence for s-states (solid circle) and p-states (solid triangle), as well as trions and biexcitons in lower panel. Sketches of the electron-hole structure of these excitonic complexes are illustrated on the right. The binding energies can be obtained from the observed optical bandgap and the quasiparticle gaps, and the results for excitonic complexes in black phosphorus\cite{xu2016extraordinarily}, WS$_2$\cite{plechinger2015identification}, MoS$_2$\cite{zhang2015valence,mak2013tightly,mai2013many}, MoSe$_2$\cite{ross2013electrical}, WSe$_2$\cite{jones2013optical,you2015observation} are presented.
 (c) Measured absorption spectrum of a single WS$_2$ layer, as sketched in the inset. Adapted with permission from Ref.\,\cite{li2014measurement}. 
\\

\textbf{Figure 4. Hybrid polaritonics.} 
(a) Schematic of the SNOM experiment performed on graphene-hBN heterostructure. (b) Experimental dispersions of hybrid plasmon-phonon-polaritons (blue circles) obtained from graphene-hBN heterostructure, compared against phonon-polariton (red triangle) from hBN. Theoretical results are shown as white lines and color map.
(c) Sketch of a charge-transfer (indirect) exciton in a van der Waals heterostructure with type-II band alignment. (d) Photoluminescence and absorption spectra of WSe$_2$ and MoS$_2$ monolayers and their hybrid heterostructure. (a-b) adapted with permission from Ref.\,\cite{dai2015graphene}, and (d) is reproduced from  Ref.\,\cite{fang2014strong}.

\bibliography{cleaned}
\onecolumngrid
\newpage~
\includepdf[pages={1-8},lastpage=8]{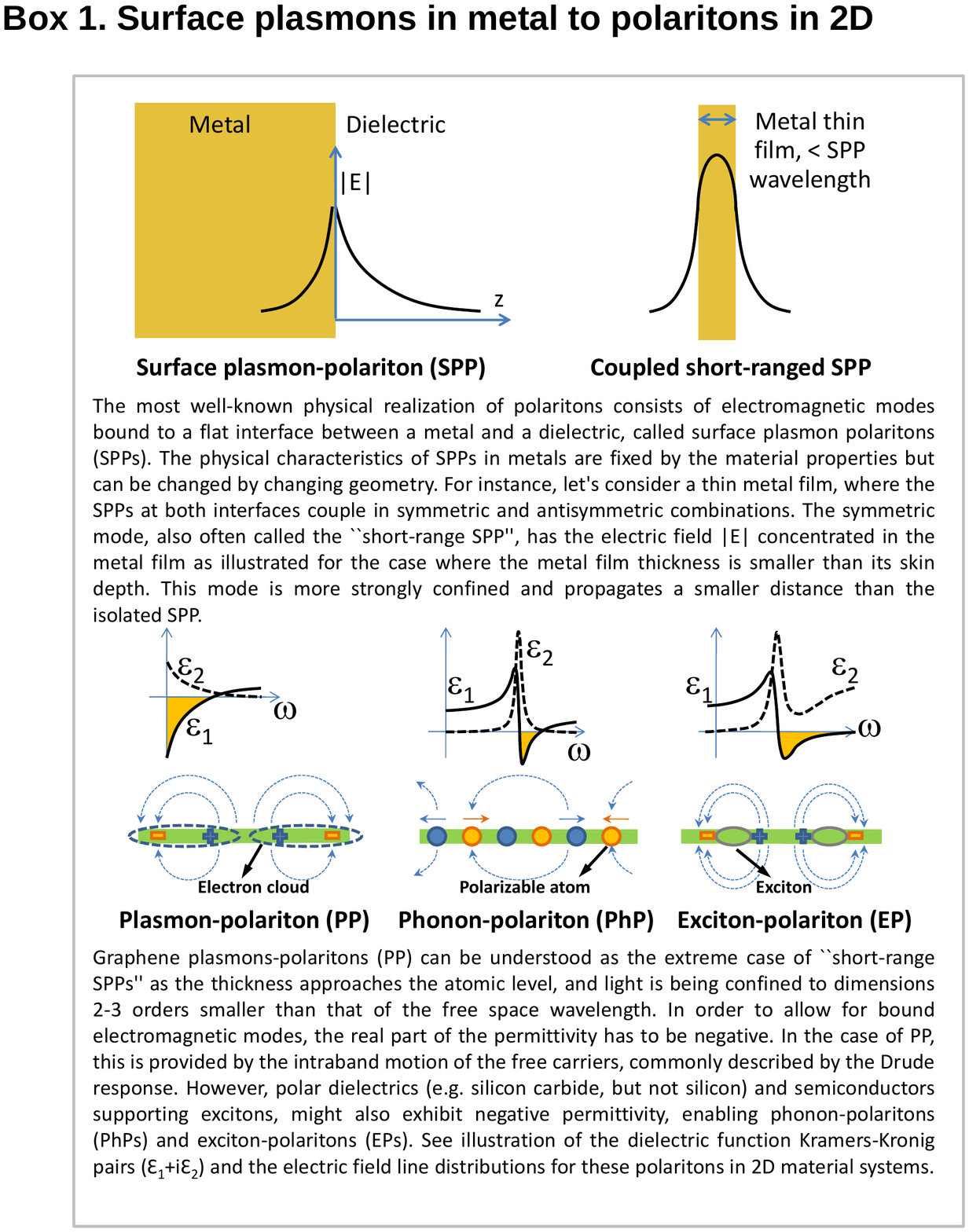}\AtBeginShipout\AtBeginShipoutDiscard 

\end{document}